\documentclass[journal]{IEEEtran}
\usepackage[pdftex]{graphicx}
\usepackage{amsmath}
\usepackage{color}
\usepackage{booktabs}
\usepackage{cite}
\usepackage[ruled,vlined,linesnumbered]{algorithm2e}

\begin{document}

\title{DER Forecast using Privacy Preserving Federated Learning}

\author{Venkatesh~Venkataramanan,~\IEEEmembership{Member,~IEEE,}
        Sridevi Kaza, Anuradha M.~Annaswamy,~\IEEEmembership{Fellow,~IEEE}
\thanks{The work reported here is based upon work supported by the Department of Energy under Award Number DOE-OE0000920. V.~Venkataramanan, S.~Kaza, and A.~M.~Annaswamy are with the Department
of Mechanical Engineering, Massachusetts Institute of Technology, Cambridge, USA e-mail: \{vvenkata,skaza7,aanna\}@mit.edu.}}

\maketitle

\begin{abstract}
With increasing penetration of Distributed Energy Resources (DERs) in grid edge including renewable generation, flexible loads, and storage, accurate prediction of distributed generation and consumption at the consumer level becomes important. However, DER prediction based on the transmission of customer level data, either repeatedly or in large amounts, is not feasible due to privacy concerns. In this paper, a distributed machine learning approach, Federated Learning, is proposed to carry out DER forecasting using a network of IoT nodes, each of which transmits a model of the consumption and generation patterns without revealing consumer data. We consider a simulation study which includes 1000 DERs, and show that our method leads to an accurate prediction of preserve consumer privacy, while still leading to an accurate forecast. We also evaluate grid-specific performance metrics such as load swings and load curtailment and show that our FL algorithm leads to satisfactory performance. Simulations are also performed on the Pecan street dataset to demonstrate the validity of the proposed approach on real data. 
\end{abstract}


\section{Introduction}

Internet-of-Things (IoT) is becoming attractive in a wide range of applications in energy, transportation, healthcare, manufacturing, and others. The rapid adoption of IoT and IoT-networks are leading to an unprecedented growth in the volumes of data that are generated by these devices. Juniper Research forecasts that the total number of IoT device endpoints will hit 83 billion by 2024~\cite{juniper}. Specifically, utilities are expected to one of the highest users, with 1.37 billion endpoints. Gartner Inc.'s report also states that \lq\lq Electricity smart metering, both residential and commercial will boost the adoption of IoT among utilities"~\cite{gartner}. Cloud computing has been proposed as a method for storing and analyzing such large volumes of data, due to the several advantages such as cost efficiency, and computing and storage capabilities \cite{six6_edge_vision}. However, a pure centralized cloud-based data storage and analytics approach becomes unrealistic due to the ever-rising data privacy concerns. General Data Protection Regulation (GDPR) in Europe lays out strict guidelines for users' data privacy, while similar laws are present in US Consumer Privacy Bill of Rights~\cite{gdpr}. Distributed methods that can efficiently and privately communicate the desired information and enable decisions are becoming more and more attractive.

In addition to privacy, other challenges that limit a centralized approach for analyzing IoT-data are the need for fast processing, low latency and sufficient bandwidth of the underlying communication network~\cite{wahabFL2021}. The cloud data centers are often deployed in locations that are far from those of data owners leading to high latency in communication, and insufficient time for real-time operation. At the same time, enabling technologies such as edge computing, wherein edge nodes such as smartphones, sensor, micro servers, autonomous vehicles and home gateways are increasingly becoming smarter~\cite{six6_edge_vision}. This in turn implies that a distributed framework with fast computing as well as fast and reliable communication to other agents in the network is becoming a feasible and viable alternative. 

The focus of this paper is the application of IoT networks in forecasting of DERs in distribution systems while ensuring privacy and security for the users. In this paper, we define DERs as electricity producing assets such as solar PV and other distributed generation (DG), and controllable loads that are capable of providing a grid function~\cite{NASEM_report}. In general, distribution system operators have limited visibility into their systems, as very few measurements are available. As DER penetration grows, this limitation becomes even more of a concern and can compromise power balance of supply and demand and therefore grid reliability. Load forecasting is an essential in distribution system operation, with long-term forecasting necessary for planning studies while mid-term and short-term load forecasting are key in day-to-day operations~\cite{one_dl_building}. However with the increase in DERs short-term load forecasting (STLF) has been proven to be a challenging task because of increased volatility. DG forecasting has also experienced similar problems, with prediction accuracy still posing challenges during operation~\cite{yadav2014DGprediction}. State-of-the-art benchmarks~\cite{three_building,four_lstm} have found that deep neural networks (DNN) are a promising solution for the STLF problem at the household level, due to their ability to capture complex and nonlinear patterns. However, neural networks require a lot of accurate and diverse training data to accurately capture these nonlinear behaviors, which is a challenge at the consumer level. 

\begin{figure}
    \centering
    \includegraphics[scale=0.65]{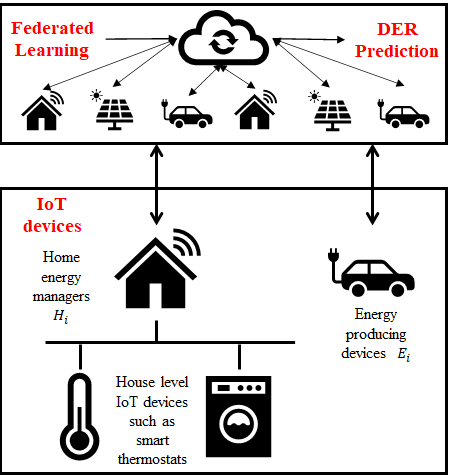}
    \caption{An overview of the DER prediction process using federated learning}
    \label{intro}
\end{figure}

The challenges of privacy as well as the requirement of large training data can be met using a distributed machine learning paradigm, Federated Learning (FL)~\cite{yang2019federated,li2020federated,lim2020federated,buildFL2020}. FL is a  machine learning framework where each device participates in training a central model without sending actual data, but only exchanges gradient information in training phase and sends prediction estimates in deployment. Current state of the art such as~\cite{six_residential_lf,seven_dl_rnn} requires that all data records are transferred from smart meters to a centralized computational infrastructure through communication networks for the training of the ML models. Consumer data is typically privately owned, and sharing of this sensitive behavioral data might have negative consequences. A solution based on FL and the use of IoT networks overcomes this hurdle of revealing consumer data to third parties. A lack of diverse training data often leads to overfitting~\cite{five_neuralnet} while deploying ML approaches. This can also be overcome using the FL method by accessing data from different nodes in a diverse IoT network such as data from several home energy managers, DGs, and Electric Vehicles. A general overview of the DER prediction process is shown in Fig.~\ref{intro}. 

The two most common issues that arise when it comes to IoT networks are privacy and security. In the context of the latter, a compelling example that has been cited in the literature is the MIRAI botnet attack~\cite{usenix2017mirai} which demonstrates that a power grid disruption can occur when a large number of IoT devices are compromised. In this paper, we focus our attention on privacy. The equally important issue of security with the use of IoT networks is not addressed in this paper. 

The contributions of this paper are as follows: (i) the development of a privacy-preserving algorithm based on Federated Learning, that allows exchange of information between IoT nodes and a global server leading to useful decisions, (ii) validation of the proposed algorithm using a simulation study of 1000 IoT nodes, with each node representing DERs such as home energy managers, electric vehicles, and solar photovoltaics, (iii) demonstration of grid services such as prediction of load swings and load curtailments based on the FL-based DER forecast, and (iv) validation of the proposed approach using real field data, Pecan street dataset~\cite{pecan}.

The overall organization of this paper is as follows. In Section II, related work is presented. The DER prediction approach using FL and neural networks is described in Section III. In Section IV, the simulation setup for validating the proposed approach is described. Simulation scenarios for demonstrating the accuracy of the proposed approach, as well as validation using real world data is presented in Section V. Finally, we wrap up with some discussion and conclusion in Section VI.

\section{Related work}
In what follows, we summarize the current state of the art grouping the related work into three different categories, DER prediction, Federated Learning, and IoT in smart grids.

\subsection{DER prediction}
Apart from forecasting at the bulk energy level and at the substation level, recent efforts of grid modernization has led to an increased focus of load forecasting closer to the consumer~\cite{six_residential_lf,hong2016review, wen2020load,kong2019lstm,kong2018tpwrs,hong2020access}. The need for load forecasting at the consumer level is driven by the increasing number of DERs, and the need for finer tuned control of resources that DER penetration necessitates. The authors of~\cite{six_residential_lf} apply a clustering technique to provides a forecast for aggregated residential load based on the practice theory of human behavior. Similar clustering techniques have been adopted for a day-ahead prediction successfully~\cite{hong2016review}. The authors of~\cite{wen2020load} focus on the load forecasting of a residential building with a one-hour resolution, while the authors of~\cite{kong2019lstm} forecast at both individual and aggregate levels with half-hourly data. Thanks to monitoring of the household appliances by separate meters, single customer forecasting improves through correlations between specific appliances~\cite{kong2018tpwrs}. In~\cite{hong2020access}, the authors show that the spatial correlations between different appliances used in a household can be leveraged to increase the accuracy of single household load forecasting. 

The studies listed above focus only on load forecasting, with a particular focus at residential level. An aggregate level forecasting has begun to be attempted only recently~\cite{khan2019smart,kobylinski2020high,smith2020privacy}. The authors of~\cite{khan2019smart} propose a forecasting model at feeder level which estimates the PV penetration, and then integrates this information into load forecasting~\cite{khan2019smart} for considering different PV penetration scenarios at the aggregate level to achieve an effective demand-side management approach. The authors of~\cite{kobylinski2020high} investigate the short-term net energy forecasting for a micro-neighborhood consisting of 75 single houses, with 15 minutes’ temporal resolution data. Reference~\cite{smith2020privacy} illustrates the use of aggregated net energy forecasting in the context of a secure energy trading platform. These studies reveal that the PV generation behind the meter increases the uncertainty, which in turn, the complexity of the net energy forecasting problem even at low-aggregate level. Much of the above literature, however, does not address the problem of user privacy when dealing with load prediction either at the consumer-level or at the aggregated level.  

\subsection{Federated learning}
Federated learning (FL) is a new machine learning paradigm that trains the ML algorithm in a distributed fashion, allowing the user data to remain local. This opens up new applications for ML algorithms, where privacy is paramount. However, several challenges still need to resolved before FL can be successfully applied to power grid problems. The reader is referred to ~\cite{yang2019federated,li2020federated,lim2020federated} for surveys on FL approaches. 
References~\cite{li2020federated} and~\cite{lim2020federated} discuss the performance of FL as compared to ML, which becomes important for critical infrastructure applications, as accuracy and implementation constraints undergo greater scrutiny for such applications.

The studies listed above provide a general overview of the federated learning process. The work in~\cite{buildFL2020,taik_2020_FL}, on the other hand,  address the specific problem as in this paper, the application of FL to load forecasting. Authors in~\cite{buildFL2020} adopt the FL framework with a well designed parameter server-client architecture, and apply this architecture for estimating HVAC performance within a building. In this paper, we utilize an architecture very similar to that in~\cite{buildFL2020} for grid-wide services such as DER forecasting and load swing prediction. We pay close attention to how the neural network is designed and how its hyperparameters are selected so as to meet the constraints of the data and the overall problem of DER prediction. Authors in~\cite{taik_2020_FL} also address load forecasting using FL, with a focus on communication efficiency of edge equipments as well as personalization of the local data at the global model so as to ensure accurate forecasting at the household level. In contrast, we focus in this paper on grid-wide needs at the distribution grid, and develop a new FL implementation, and show how the FL algorithm can be trained so as to ensure privacy of consumer data and at the same time lead not only to accurate DER forecasts but also desired grid-services. 

\subsection{IoT device application in smart grids}

IoT devices are experiencing an exponential growth in all sectors including transportation and smart cities~\cite{palaniswami_iot}. With much of the innovation in smart grids occurring at the grid edge, IoT is poised to play a major role in enabling dynamic power balance at all points of the grid, especially at the distribution grid level. Much of the existing literature focuses on applications of IoT devices to smart homes, demand response, and related smart communities based applications~\cite{home_iot,flisr_iot,yuxin2010iot,siano2010iotsmartcity,iot_power_venayagamoorthy}. IoT devices have already become ubiquitous at the home level, with the proliferation of smart home devices such as Nest thermostats, and platforms such as Google Home and Alexa~\cite{home_iot}. IoT devices have been proposed in the management of extended outage conditions, such as automated fault location, isolating and service restoration (FLISR) in~\cite{flisr_iot}. IoT devices also provide a way to determine the position of the defective parts, separates them, and applies switching task to recover the largest number of healthy part of the affected energy feeder by having increased sensors deployed in control devices~\cite{yuxin2010iot}. Also, at the advanced level, this function can be developed by using self-healing methods that are able to activate the participation of the customers as well as of dispersed generation units~\cite{yuxin2010iot}. Various real-world deployments detail the benefit of increased sensing using IoT devices for various applications ranging from traffic management to urban innovation~\cite{siano2010iotsmartcity}. Implementing these strategies leads to increase the reliability, power quality and profits~\cite{iot_power_venayagamoorthy}. Further applications of IoT technology in the power system domain can be found in the survey~\cite{iot_power_venayagamoorthy}.

In the specific context of this paper relating to accurate DER forecast, IoT technology is poised to play a major role. The two major hurdles in its implementation are privacy and security. This paper pertains to the first, and proposes the use of Federated Learning so as to ensure privacy preservation and leverage the IoT technology to lead to accurate DER forecasts.

\section{DER prediction using Federated Learning}

We begin with an overview of the neural network procedure to be utilized in FL. The underlying problem is a nonlinear mapping $f(\cdot)$ between a vector of inputs $x$ and an output $y$. A neural network is an extraordinarily effective tool in learning this mapping, and constitutes a bulk of the ML approaches used for learning~\cite{yadav2014DGprediction,four_lstm,kong2019lstm}. A typical process by which a neural network learns the nonlinearity $f(\cdot)$ is through a training and testing phase. 

The training phase consists of a vector of inputs $x_t$ collected at epoch $t$, each of which leads to an output $y_t$. The total number of samples in epoch $t$ is defined as $n$. A typical neural network architecture, referred to as a deep-learning network consists of multiple layers, interspersed with activation function with the output $y_t$ related to $x_t$ in the form indicated in~\eqref{dnn}. 

\begin{equation}
    y_t = \sum_{i=1}^N w_{2i}^t\phi (w_{1i}^y x_t + b_i)
    \label{dnn}
\end{equation}

\noindent where $\phi(\cdot)$ denotes an activation function, $N$ is the number of neurons, and $w_{ji}$ are the weights of layer $j$, $b_i$ is a bias term, and represents the input-output relation for a network with one hidden layer. The training procedure then consists of adjusting the weights $w_{ji}$ as
\begin{equation}
    w_{ji}^{t+1}=w_{ji}^{t}+\delta^t
    \label{training}
\end{equation}

\noindent where $\delta^t$ corresponds to the gradient of a loss function $L$ which is defined as,

\begin{equation} 
    \delta^t=~-\eta\nabla L^t|_{w^t} 
    \label{obj}
\end{equation}

\noindent where $\eta$ is a learning rate, and $w^t$ is a vector of all weights $w_{ji}^t$ at an iteration at epoch $t$. A typical loss function $L$ is a mean squared prediction error over $n$ samples which is defined as,

\begin{equation}
    L^t = \frac{1}{n} \sum_{i=1}^n (Y_i - y_i)^2
    \label{error}
\end{equation}

\noindent Through a repeated training of these weights, the output $y_t$ of the neural network is then allowed to approximate the true value, $Y_t$. The testing phase then consists of using a distinct set of $x$, which is different from the training phase, and using the trained neural network to predict the corresponding output $y$. 

One can use the above neural network to forecast the power consumption (or generation) of a device $d$ (or a renewable generation source) in the following manner. Suppose $y$ denotes $\widehat P_d(T)$, the power-consumption forecast at time $T$, and the input $x$ denotes a vector $\{P_d(T-n_1),P_d(T-n_2),\ldots P_d(T-n_M)\}$, where $T-n_i$ corresponds to a past instant relative to $T$; that is, the power consumption on a given day at a given hour may be correlated with power consumptions at the same hour during the previous day, the power consumption a few minutes before $T$, or a combination thereof. The underlying relation between such a vector of previous consumptions before $T$, $x$, and the consumption $y$ at time $T$ can be viewed as a nonlinear relation $f(\cdot)$. The neural network as in~\eqref{dnn} is tasked with learning this mapping using the training procedure using which the weights $w^t$ are trained using several samples until the loss function $L^t$ falls below a threshold $\epsilon$.  That the neural network has indeed been satisfactorily trained is tested by fixing the neural network with the converged weights and evaluating the loss function for a distinct set of inputs $x^t$ not utilized for training and testing using the procedure as described above. We shall refer to this overall neural network as a global model $G$. That is, the global model $G$ can perform the computations in (\ref{dnn})-(\ref{obj}) repeatedly to forecast the power consumption of a device $d$. This is the typical procedure using in any ML that uses neural networks.

It should however be noted that the above procedure implies that data from device $d$ in the form of $x$ has to be sent to a global model repeatedly during the training process. That is, every device $d$ will then have to share the data $\{P_d(T-n_1),P_d(T-n_2),\ldots P_d(T-n_M)\}$ for every $T$ and every device $d$ repeatedly. Given the pervasive and large number of IoT devices that generate this data, it is unrealistic to expect all users to consent to their data being accessed to create machine learning models in such a manner, especially if the devices are not owned by the utility. For this purpose, we introduce a variation in the ML training process, using the tenets of FL~\cite{yang2019federated}, and is described below.

\begin{figure}[htbp]
    \centering
    \includegraphics[scale=0.45]{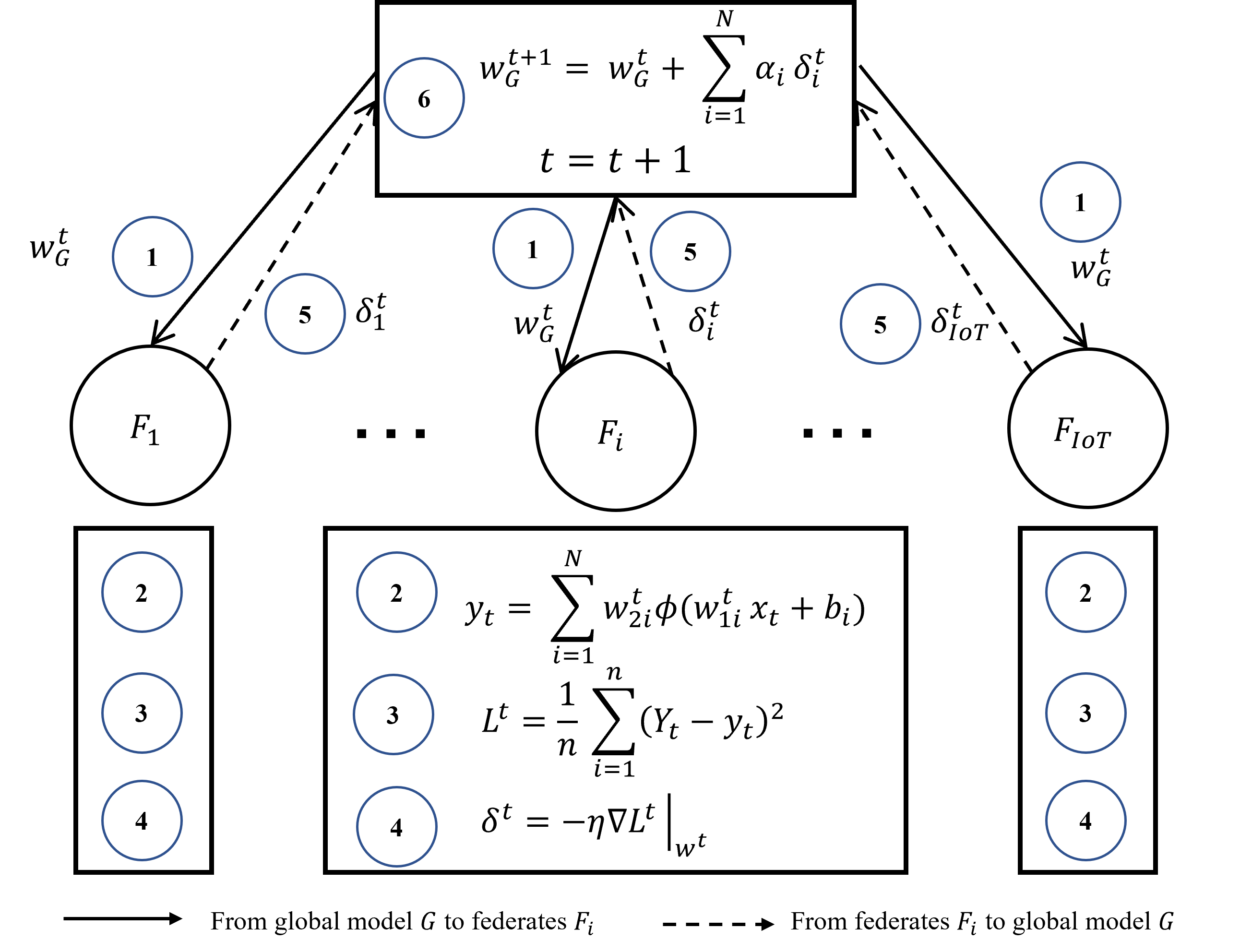}
    \caption{The schematic of neural network training using federated learning is shown here. The steps (1)-(6) are repeated until $L^t \leq \epsilon$ }
    \label{block}
\end{figure}

In what follows we assume that all IoT devices under consideration can be grouped into two types, $H_i$ and $E_i$, where $H_i$ corresponds to home energy managers that may manage a collection of home appliances including HVAC, smart thermostats, smart refrigerators, dishwashers, laundry machines, and other home appliances. Devices $E_i$ correspond to rooftop solar panels, EVs, and storage units. Both types of devices are assumed to be connected to the secondary feeder network, with the total number of devices given by $N_H$ and $N_E$, with $N_H+N_E=N_{IoT}$. The power consumed by $H_i$ and $E_i$ at time $T$ is given by $P_{H_i}(T)$ and $P_{E_i}(T)$. We refer to each of these devices $H_i$ and $E_i$ as a federate $F_i$ for the FL process. During the training process, the adjustment of the weights $w^t$ of the neural networks proceeds described in (\ref{dnn})-(\ref{obj}) in the following manner. 

The global model sends initial weights $w_G^t$ at time $t=0$. The local model then generates several samples $x_t$ that corresponds to a range of time-instants $T$ as well as the resulting true output $Y_t$, and the predicted $y_t$ for the weights $w^t_G$ using (1). The corresponding loss function $L^t$ is then computed by the local model as in~\eqref{error} and the gradient $\delta^t$ as in~\eqref{obj}. The local model, which corresponds to federate $F_i$, then sends this gradient $\delta_i^t$ to the global model $G$. The global model then updates its weights $w_G^t$ as $w_G^{t+1}$ using the collection of all gradients from the federates $F_i, i=1,\ldots, N_{IoT}$ as,

\begin{equation}
    w_G^{t+1} = w_G^t + \sum_{i=1}^{N_{IoT}} \alpha_i \delta_i^{t}
    \label{fl}
\end{equation}

\noindent where $\alpha_i$ is a pre-determined set of weights that combines all gradients. The global model then sends the updated weight $w_G^t$ back to each of federates $F_i$. The federate $F_i$ in turn collects new input-output pairs $(x_t,y_t)$ using the updated weights $w^t$ using~\eqref{dnn}, computes the new loss function $L^t$ using~\eqref{error}, and the new gradient using~\eqref{obj}. This new gradient is then sent by the federate $F_i$ to the global model $w_G$, and the whole training process repeats.  The output $y_t$ of the neural network corresponds to the predicted outputs $\widehat P_{H_i}(T)$ or $\widehat P_{E_i}(T)$ for federate $F_i$, at time $T$. The overall schematic of the neural network training using FL is presented in Fig.~\ref{block}.

It should be noted that throughout the training process, each local model only sends the gradient $\delta_i^t$ to the global model, does not need to send $x_t$. As a result, the actual data $x_t$, $Y_t$, and the predicted data $y_t$ stays local to the federate $F_i$. The only information revealed to the global model is $\delta^t$ and therefore preserves privacy of the device $F_i$. In addition, the FL procedure described above is such that the global model is able to utilize a large variety of federates $i=1,\ldots N$, and therefore leverage a rich variety of training data leading to be better prediction. Finally, our procedure is applicable in equal measure to both IoT devices of type $H_i$ and $E_i$, and therefore includes both distributed generation and distributed loads.

\begin{algorithm}[htp]
  \SetAlgoLined
  \SetKwFunction{algo}{FL}\SetKwFunction{proc}{LocalTraining}
  \SetKwProg{myalg}{Algorithm}{}{FL}
  \myalg{\algo{}}{
   Set $t~=~0$\;
   For $t \leq 150$\;
   Initialize weights $w_G^t$ at the global model $G$\;
   Send $w_G^t$ to each federate $F_i$\;
  \proc{$F_i, w_G^t$}\;
   Set $\delta_i^t =\delta^t$\;
   Compute $w_G^{t+1}$ using (5)\;
   Set $t=t+1$\;
   Repeat until $L^t < \epsilon$, a pre-determined tolerance\;
  \KwRet\;}{}
  \SetKwProg{myproc}{Procedure}{}{}
  \myproc{\proc{$F_i, w_G^t$}}{
  Set $w^t=w_G^t$ in (1)\;
  $F_i \gets (x_t,y_t)$ of $n$ samples using (1)\;
  Compute $L^t$ using (4)\;
  Compute $\delta^t$ using (3)\;
  \KwRet $\delta_i^t$\;}
  \caption{Federated Learning Algorithm}
\end{algorithm}

A typical process of DER-forecast can occur in the following manner. Collect the input-output pair $[x_t, \hat{P}^t(T)]$ for a federate $F_i$ for several samples $n$. An example of $x_t=[P^t (T-15), P^t (T-30), P^t (T-60), P^t (T-120), P^t (T-1440)]$, where $P^t(T-m)$ denotes the actual power consumption, and $m$ denotes the minutes prior to time $T$. The number of samples $n$=2880, obtained by collecting data every 15 minutes over a period of 30 days. The overall training procedure of the FL-based neural network is summarized in Algorithm 1. 

A more simplified training procedure can be adopted, compared to Algorithm 1, and is utilized in the results reported in the subsequent sections. This is briefly described here. The total number of samples available each day for training is 96. Rather than use 96 samples from all 30 days, a day was chosen at random, and the computations at each Federate were carried out using the mini-batch of samples from that day. A tolerance of $\epsilon=0.001$ was chosen. A total of 150 epochs was found to be sufficient to achieve this desired tolerance. The neural network consisted of 2 hidden layers with 20 neurons in each layer. All of these hyperparameters of the simplified training procedure are shown in Table 1. 
 
\begin{table}
    \centering 
    \normalsize
    \caption{Neural network hyper-parameters}
    \label{nnhyper}
    \begin{tabular}{cc}\toprule
    Hyper-parameter & Result\\ \midrule
    Mini-batch size $n_b$    & 96 samples \\
    Learning rate  $\eta$     &  0.001\\
    Maximum epoch $n_e$     & 150\\ \bottomrule
    \end{tabular}
\end{table}

In the testing phase, the model performance is evaluated by testing with a new set of data from the next month, from a similar season to ensure that the model is still valid. In the case studies, we will use the root mean square error (RMSE), which coincides with the square root of the loss function $L$ defined in~\eqref{error} to quantify the performance of the FL algorithm. That is,
\begin{equation}
    \text{RMSE}^t=\sqrt{L^t}
    \label{rmse}
\end{equation}

\section{Simulation setup}
To validate the concept of federated learning for privacy preserving load prediction, we create a numerical testbed that comprises of three components - (1) Power physics simulation, (2) Federated learning platform, and (3) Grid service model. 

\subsection{Power physics simulation}
The physical layer simulation needs to generate the IoT level data corresponding to the various grid components. Various physics simulators such as building models, GridLAB-D, and others can be used for this purpose. GridLAB-D ~\cite{gridlabd} provides detailed models for various power system components, with active ongoing upgrades to existing models. GridLAB-D uses an advanced algorithm to solve the power system simultaneously by solving the states for all the different devices at the same time, and not sequentially and therefore offers the flexibility to develop complex models, and implement user developed algorithms for various control purposes. In the context of the current problem, we will utilize GridLAB-D to generate physical data from the distribution system ranging from the primary feeders and secondary feeders, all the way to consumer buildings, and individual IoT devices such as HVAC units, EVs, and rooftop PVs. This makes the platform suitable to validate the overall forecast using the FL Algorithm 1.

\begin{figure}
    \centering
    \includegraphics[scale=0.6]{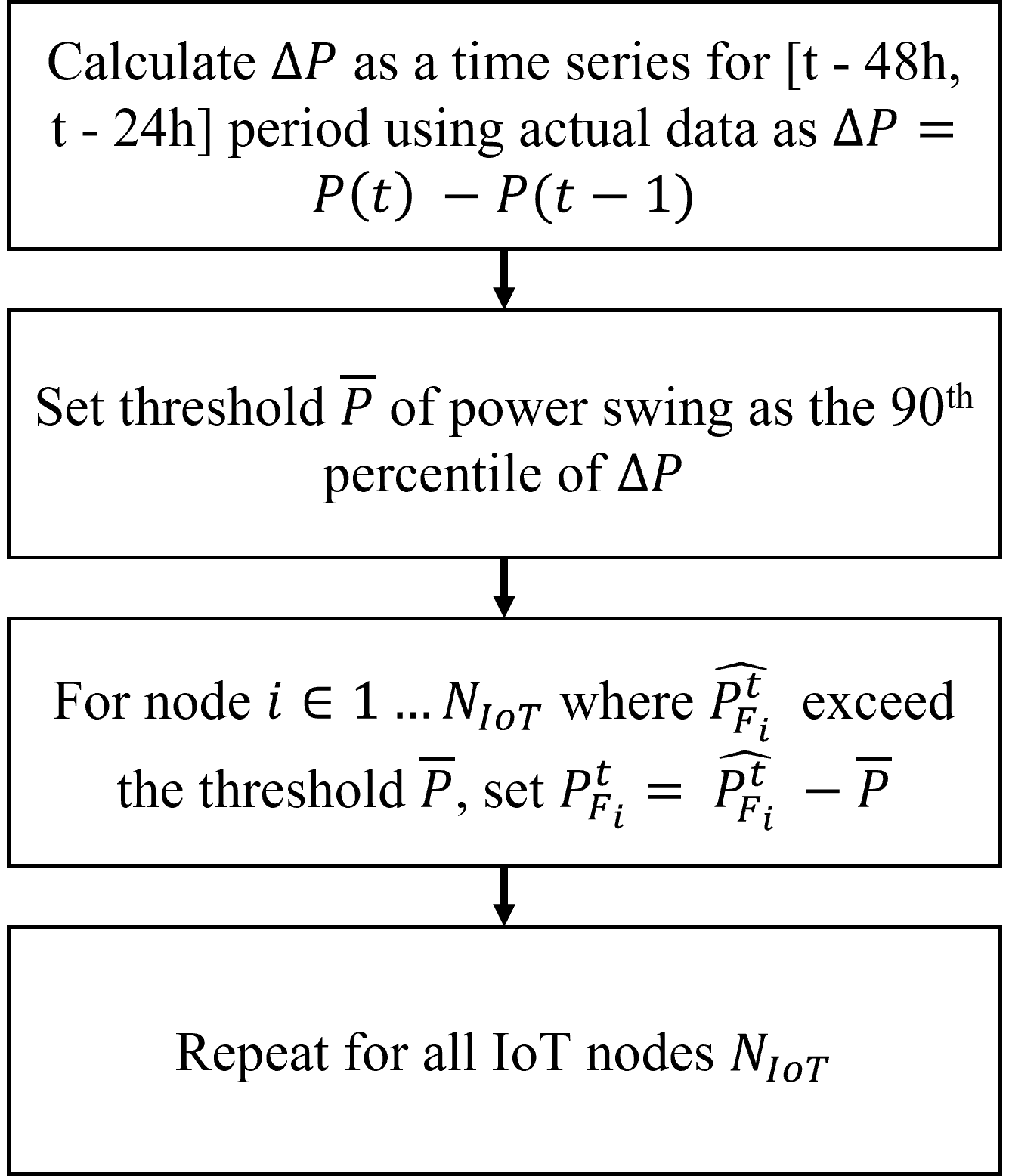}
    \caption{A simple peak shaving algorithm to minimize power swings}
    \label{peak_shaving}
\end{figure}

\subsection{Federated Learning platform}
\begin{figure*}
    \centering
    \includegraphics[scale=0.68]{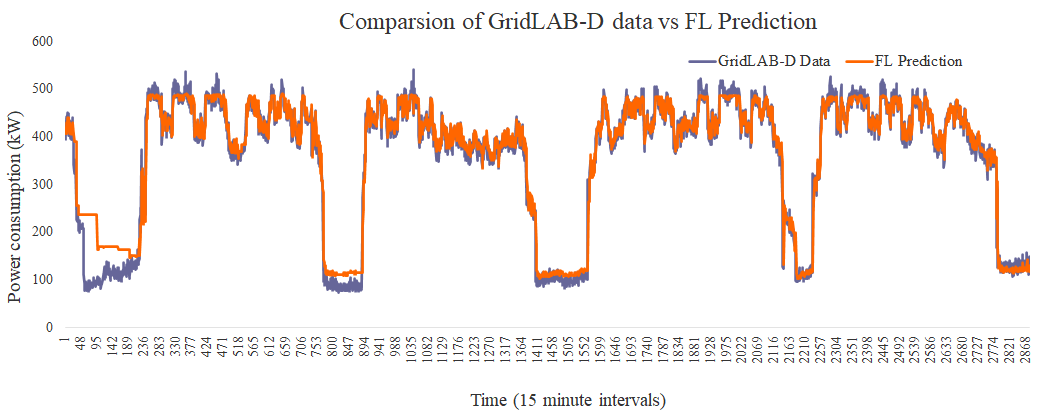}
    \caption{Load prediction using Federated learning}
    \label{load_pred}
\end{figure*}

To implement federated learning, we leverage the approach in ~\cite{buildFL2020} denoted as the Building Federated Learning (BuildFL) platform. BuildFL adopts the popular parameter server (PS) architecture, in which a server is created to store the parameters of the ML model (in this case, the weights $w_G^t$ of the neural network) and then serves them to the clients, which are the federates $F_i$. The federates creates a local update by training with the data set it collects and exchanges model parameters to the global model $G$. In BuildFL, building sites are workers and a cloud server is created to serve as the parameter server. The BuildFL platform integrates machine learning libraries such as PyTorch, which is mainly used for computer vision or natural language processing, and its library only supports gradient-based models. To support other models that facilitate building analytics, such as bagging and boosting algorithms, BuildFL integrates Scikit-Learn, which includes the library of models such as random forest, boosting tree, etc. 

BuildFL provides a uniform function template for distributed training of different models. As detailed in Section II, a DNN with two hidden layers was chosen as the ML architecture in this paper. The  hyper-parameters of the neural network have been defined in Table~\ref{nnhyper} and stochastic gradient descent (SGD)~\cite{bottou2012sgd} is chosen as the optimization algorithm to update the weights.

\subsection{Grid service model}

In order to demonstrate the impact of accurate DER forecast using our proposed FL method, we carry out a particular grid service using the forecast, which corresponds to predicting load swings and curtailing them using a peak shaving algorithm. Such a load swing prediction is of growing importance with increased DER penetration, as evidenced by the focus on the "duck curve" problem~\cite{duckcurve2015nrel}. This problem corresponds to the large unserved load and its increase over a short period of time as solar based resources disappear, which occurs on a daily basis at dusk. We assume that a certain percentage of the loads are flexible and directly controlled by the utility, and that once these loads are accurately forecasted, then they can be commanded to follow a particular command signal. One such algorithm is described in Fig. 3, and we will utilize this to  demonstrate the impact of the FL-based DER forecast described in Section II.

\begin{figure*}
    \centering
    \includegraphics[scale=0.55]{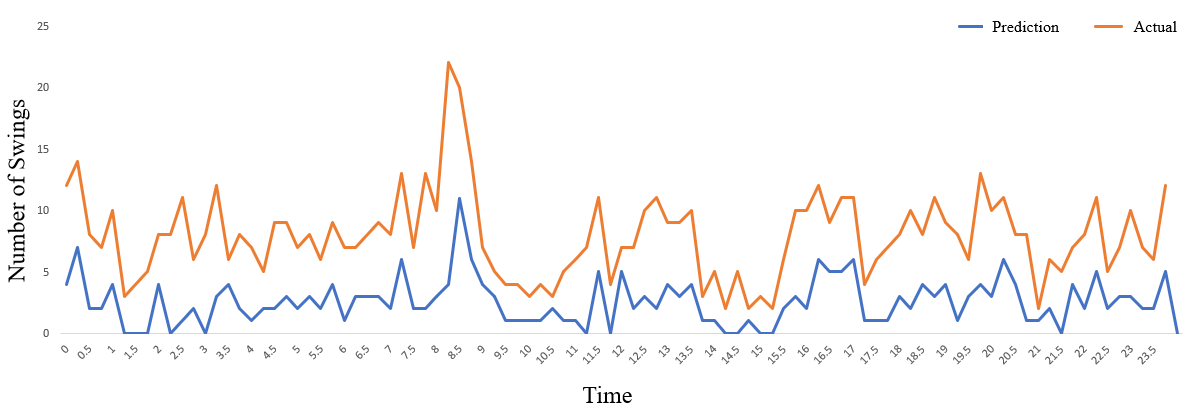}
    \caption{Time of maximum power swings}
    \label{time_ps}
\end{figure*}

The details of the load swing prediction proceed as follows. First, a threshold value $P_T$ is computed by generating a time-series $\Delta P_t$, which corresponds to power consumption change in $P$ over $[t-1, t]$, and determining the 90th percentile of $\Delta P_t$ over a 24-hour period. A load swing is said to occur at $T_0$ if $\Delta P_{T_0} > P_T$. One can therefore predict  load swings using the power consumption forecast that is carried out using the FL method described above by computing $\Delta \widehat P_t$ and determining load swings using $P_T$. Once these load swings are predicted, the next step in the grid-service model is to command a flexible load to be curtailed as described in Fig. 3. In the next section, we will demonstrate this grid service using data obtained from 1000 nodes.

\begin{figure*}[htbp]
    \centering
    \includegraphics[scale=0.58]{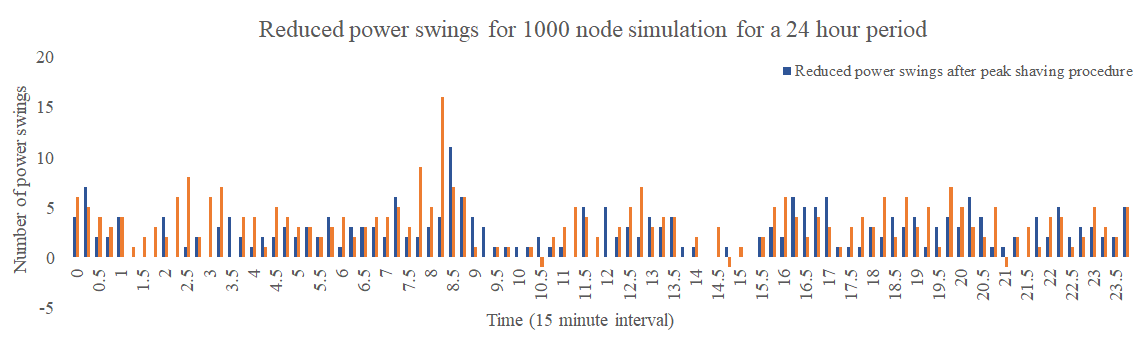}
    \caption{Reduction in power swings in consecutive days}
    \label{reduction_ps}
\end{figure*}

\section{Use-cases}
 
In this section, we validate the proposed FL-method for DER forecast using a use-case of a distribution feeder with 1000 nodes. The goal is to (i) predict over a (a) 24-hour period the power consumption at time $T$ of 1000 nodes at 15 minutes prior to $T$, i.e., at $T-15$ for all 1000 nodes (b) the number of load swings of these nodes at also $T-15$, and (ii) curtail these load swings at some of these nodes using a peak shaving procedure detailed in Fig.~\ref{peak_shaving}. GridLAB-D is used to simulate the distribution feeder with 1000 nodes, each of one was modeled to emulate a distribution feeder node. The GridLAB-D model was built as one distribution feeder with 1000 nodes, and the load shapes were synthetically generated. We accomplished this by starting from a single default load profile from GridLAB-D, and created the other 999 load profiles synthetically using a standard Gaussian distribution (zero mean, $\sigma = 0.1$) around this single load profile.  Algorithm 1 was then employed to predict the power consumption of each of those nodes. The input vector was chosen to be a six-dimensional vector as in Section II, with $n_1$~=~15min, $n_2$~=~30min, $n_3$~=~60min, $n_4$~=~90min, $n_5$~=~120min, and $n_6$~=~1440min.

The results from the load prediction and the peak shaving procedure are presented below. Finally, an evaluation of this method is carried out using a realistic dataset from the Pecan Street Inc. Dataport~\cite{pecan}.

\subsection{Prediction of load}

We illustrate in Fig.~\ref{load_pred} the prediction of the power consumption of a node over a 24-hour period. This prediction (shown in orange) is compared with the actual consumption (shown in blue). The prediction and data correspond to a node picked at random out of 1000 nodes. The training occurred over several epochs of data collected every 15 minutes over a 30 day period. The results illustrate the accuracy of the FL approach in estimating the power consumption by the nodes. The prediction accuracy is observed to improve with larger amounts of training data, and sensitive to the hyper-parameters. Weekends are observed to have a significant difference over weekday consumption, which is a characteristic of the nature of the chosen distribution feeder. An RMSE was calculated as in~\eqref{rmse} and averaged for all the 1000 nodes and was found to be 1.642, validating the accuracy of the proposed FL method. 

We illustrate the usefulness of the power consumption prediction using a second performance metric. For this purpose, we first plot the average number of load swings for the 1000 nodes over a 30 day period in Fig.~\ref{time_ps}, with a load swing as defined above in Section III. The X-axis represents the time of day at which the load swings happen over a 24 hour period, and the Y-axis represents the average number of swings that occur at that time over a month. The orange curve represents the actual number of load swings determined from the GridLAB-D data, while the blue curve represents the predicted load swings from the FL procedure. It is observed from the figure that the FL procedure is able to predict the time of swings accurately, as evidenced by the peaks occuring at the same time in both curves, even though with an under-prediction in the magnitude. More sophisticated procedures for defining load swings may help improve this prediction. 

\begin{figure*}
    \centering
    \includegraphics[scale=0.64]{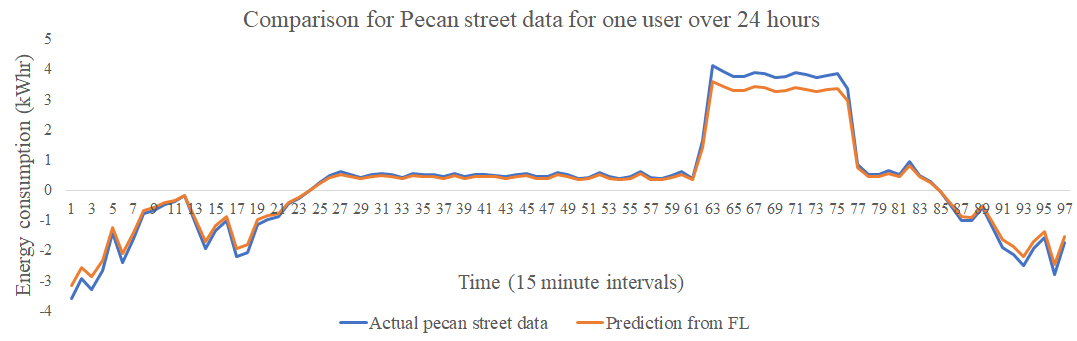}
    \caption{Comparison of prediction and actual data from Pecan street dataset for a week long period}
    \label{pecan_compare}
\end{figure*}

\subsection{Reduction in power swings}

We now utilize the load-swing prediction to carry out a load-curtailment, assuming that flexible loads that are compatible with Demand Response are present among the 1000 nodes. The threshold $P_T$ was set to 31kW based on the 90th percentile of all power consumption over the 1000 nodes over a 2-month period. When at $T-24h$, if it was observed that $\widehat P(T)> 31kW$ at a particular node, a direct load control signal was sent to curtail the power consumption at that node by enforcing a reduction of 31kW, as indicated in Fig.~\ref{peak_shaving}. Figure~\ref{reduction_ps} illustrates the result of such a peak-shaving algorithm over a 24 hour period. The orange bar represents the power swing without any load curtailment, while the blue bar represents the load swings after peak shaving. While the results indicate an overall reduction in the number of load swings on most days, it clearly illustrates that the power prediction is not able to mitigate all swings from happening. The negative ticks are also a result of an over-prediction of the power swings.

We have made an assumption here that a peak shaving algorithm as in Fig.~\ref{peak_shaving} can be implemented, in deriving a grid performance as shown in Fig.~\ref{reduction_ps}. This requires not only the loads to be flexible but also curtailable when demanded by the utility. Such a direct load control may not always be possible and may require a transactive control architecture with suitable incentives or rebate programs~\cite{annaswamy2015transactive}. 

\subsection{Pecan street validation}

The FL procedure is also tested on real world data from Pecan street. Pecan street is a project that provides data of household consumption and DER behavior from various locations such as Austin~\cite{pecan}. As in the earlier Gridlab-D exercise, here too the neural network is trained for using  data  from a single building collected specifically over 30 weekdays, and then the trained model is used to predict the load for a random weekday in the next month. Fig.~\ref{pecan_compare} shows the comparison between the actual load consumption and the prediction, and the RMSE value for the model's prediction is 1.98, once again demonstrating the accuracy obtainable using the proposed FL method. A similar procedure can be adopted for predicting power consumption over weekends by suitably collecting training data over weekends rather than weekdays, over a period of time. 
\subsection{Discussion}

It is important to note the effect of selecting the right features on the accuracy of the trained neural network model~\cite{Mu2010STLFSimilarDays,Din2017STLFDNN,he2017load}. The underlying data associated with the DER demonstrates a significant periodicity, which is evident in Fig.~\ref{load_pred} and Fig.~\ref{reduction_ps}. It is important to pick appropriate features of the historical data to ensure that the model estimates the non-linearities properly. Authors in~\cite{Mu2010STLFSimilarDays} have made such an observation that a choice of similar days for training drastically improves the forecast accuracy. Previous studies have shown that filtering out weekends, public holidays, and other out-of-normal consumption patterns can also significantly improve prediction accuracy~\cite{he2017load}. In the Pecan street experiment in particular, we have only focused on weekdays, and filtered out other holidays from the data. When such a filtering was not included, the prediction error was observed to significantly increases, to as much as 30\%. In addition to historical data, other features can also be used to improve the prediction accuracy, such as weather, ambient temperature, wind speed, and price of electricity (especially with demand response). 

The accuracy of the FL approach has been observed to be somewhat less accurate than the traditional centralized ML one~\cite{li2020federated,buildFL2020}. We have not carried out a comparison of the proposed FL approach with a general ML in this paper, as our objective was to demonstrate a DER prediction mechanism that ensures user's privacy. Further research and experimentation is required to  better explore the trade-offs between privacy and accuracy in an FL method.



\section{ Summary and Concluding Remarks}
In this paper, we have considered the problem of DER prediction in a distribution grid at the consumer  level using IoT nodes. Each IoT node is assumed to represent DERs such as renewable resources (ex. solar PV), storage such as electric vehicles, and home energy managers consisting of an aggregated set of flexible loads such as HVAC and thermostatically controlled appliances. The problem of accurate DER prediction is of paramount importance in distribution grids, not only because distribution systems have limited sensing capability but also because they can directly impact grid-specific performance such as power balance. The challenge, however, is that typical approaches used for DER forecast are data-centric and require consumers to share their consumption/generation data that can compromise their privacy. The contribution of this paper is a distributed algorithm that is based on Federated Learning which transmits a model of the consumption and generation patterns without revealing consumer data. 
We have described in this paper this privacy-preserving algorithm in detail, and carry out its validation using a simulation study of 1000 IoT nodes, leading to a forecast with an RMSE of  1.3. We have also demonstrated grid services such as prediction of load swings and load curtailments based on the forecast. Finally, we validated the proposed approach using real field data, with an RMSE of 1.98. Future research will involve a more detailed incorporation of grid physics such as optimal power flow and DER constraints such as adjustable loads,  communication efficiencies of FL, and market structures with transactive energy and related incentives.

\bibliographystyle{IEEEtran}
\bibliography{refs}

\end{document}